\documentclass[conference,letterpaper]{IEEEtran}

\usepackage[utf8]{inputenc} 

\usepackage[T1]{fontenc}
\usepackage{url}
\usepackage{ifthen}
\usepackage{cite}
\usepackage[cmex10]{amsmath} 

\usepackage{amssymb}
\usepackage{graphicx}
\usepackage{url}
\usepackage{epsfig}
\usepackage{graphicx}
\usepackage{amsfonts}
\usepackage{amssymb}
\usepackage{amsbsy}
\usepackage{amsfonts}
\usepackage{amsthm, mathrsfs}
\usepackage{longtable}
\usepackage{stackengine}
\usepackage{mathtools}
\usepackage{stmaryrd}
\usepackage{relsize}
\usepackage{multirow}
\usepackage{bm}
\usepackage[T1]{fontenc}
\usepackage{url}
\usepackage{subcaption}
\usepackage{caption}
\usepackage{dsfont}
\usepackage{xfrac}
\usepackage{color}                    
\usepackage{xfrac}
 
\usepackage{setspace}

\usepackage[T1]{fontenc}         
\usepackage{amstext}
\usepackage{amssymb,amsmath,amsthm,enumitem}
\usepackage{sansmath}

\usepackage{mathrsfs}
\DeclareMathAlphabet{\mathpzc}{OT1}{pzc}{m}{it}

\theoremstyle{definition}

\newtheorem{definition}{Definition}

\usepackage{stfloats}

\def\markov{\hbox{$\--$}\kern-1.5pt\hbox{$\circ$}\kern-1.5pt\hbox{$\--$}}

\makeatletter
\newcommand*{\centernot}{%
  \mathpalette\@centernot
}
\def\@centernot#1#2{%
  \mathrel{%
    \rlap{%
      \settowidth\dimen@{$\m@th#1{#2}$}%
      \kern.5\dimen@
      \settowidth\dimen@{$\m@th#1=$}%
      \kern-.5\dimen@
      $\m@th#1\not$%
    }%
    {#2}%
  }%
}
\makeatother

\usepackage{cite}
\usepackage{amsmath,amssymb,amsfonts}
\usepackage{algorithm}
\usepackage{algpseudocode}
\usepackage{graphicx}
\usepackage{textcomp}

\usepackage[table]{xcolor}  
\usepackage{color, colortbl}

\usepackage{hyperref}
\hypersetup{
    colorlinks=true,
}

\begin{document}
\title{Privacy-Preserving Near Neighbor Search via\\Sparse Coding with Ambiguation} 


 \author{%
   \IEEEauthorblockN{Behrooz~Razeghi\IEEEauthorrefmark{1},
                     Sohrab~Ferdowsi\IEEEauthorrefmark{2},
                     Dimche~Kostadinov\IEEEauthorrefmark{3},
   				     Flavio.~P.~Calmon\IEEEauthorrefmark{4},
                     Slava~Voloshynovskiy\IEEEauthorrefmark{1}
                     }
    \IEEEauthorblockA{\IEEEauthorrefmark{1}%
                     University of Geneva
                   }
    \IEEEauthorblockA{\IEEEauthorrefmark{2}%
                     HES-SO Geneva}
    \IEEEauthorblockA{\IEEEauthorrefmark{3}%
                     University of Zurich}       
    \IEEEauthorblockA{\IEEEauthorrefmark{4}%
                     Harvard University}
                     }

\maketitle

\begin{abstract}
In this paper, we propose a framework for privacy-preserving approximate near neighbor search via stochastic sparsifying encoding. 
The core of the framework relies on sparse coding with ambiguation (SCA) mechanism that introduces the notion of inherent shared secrecy based on the support intersection of sparse codes. 
This approach is `fairness-aware', in the sense that any point in the neighborhood has an equiprobable chance to be chosen. 
Our approach can be applied to raw data, latent representation of autoencoders, and aggregated local descriptors.
The proposed method is tested on both synthetic i.i.d data and real large-scale image databases. 
%
%

\end{abstract}




\vspace{2pt}

\section{Introduction}
\label{Sec:intro}

\vspace{-1pt}


%
Many modern signal processing, machine learning and data mining applications, such as biometric authentication/identification, pattern recognition, speech processing and recommender systems, require near neighbor search of a query with respect to a given dataset, and a distance measure. 
Many search services are outsourced to third parties (service providers) who possess powerful storage, communications and computing facilities.  
The major challenge is to satisfy privacy constraints of the owner's data and the clients' interests, while still being capable of performing the fast search service in multi-billion entry datasets.

Let $\left( \mathcal{S}, d_{\mathcal{S}} \right)$ be a metric space. Given a set  $\mathcal{X} \subseteq \mathcal{S}$ of $M$ points, a parameter $r$, and a query point $\mathbf{y} \in \mathcal{S}$, the goal of the exact near neighbor (NN) problem is to find a point $\mathbf{x} \in \mathcal{X}$ such that $d_{\mathcal{S}} \left( \mathbf{x}, \mathbf{y} \right) \leq r$, if such a point exists. 
In the approximate variant of this problem (ANN), given $c > 1$, the problem is relaxed to find a point $x \in \mathcal{X}$ such that $d_{\mathcal{S}} \left( \mathbf{x}, \mathbf{y} \right) \leq cr$. 
Theses problems can be generalized to $k$-NN and $k$-ANN setting. 
%
In this context, we assume $\mathcal{S}$ to be the $N$-dimensional Euclidean space, i.e., $\mathcal{S} = \mathbb{R}^N$, and the distance given by an $\ell_2$-norm, $d_{\mathcal{S}} \left( \mathbf{x}, \mathbf{y}\right) = \Vert \mathbf{x} - \mathbf{y}\Vert_2$. 
The NN search based on na\"{i}ve solution, i.e., the linear scan, is the bottleneck of the system in large scale high-dimensional data sets \cite{wang2017survey}. 
Alternatively, approximate near neigbor (ANN) search, is more efficient in terms of query time and space complexity \cite{indyk1998approximate, gionis1999similarity, wang2015learning}. 
Perhaps the most popular solution to ANN problem is via hashing, 
where the aim is to \textit{transform} the data points to a lower dimensional space, then perform similarity search in the lower dimensional representation.  
%
%
%
The two main research directions are 
(1) 
Locality Sensitive Hashing (LSH): 
indexing points using a hash table with the property that similar (closer) data points have a higher probability 
of collision than dissimilar (far) points \cite{wang2017survey, indyk1998approximate, gionis1999similarity, wang2015learning, datar2004locality,  christiani2019fast, har2019near}; 
(2) 
learning to hash: 
performing NN similarity search in a low dimensional space with a lower search complexity \cite{salakhutdinov2009semantic, weiss2009spectral, Sohrab_ISIT2017, wang2017survey, Razeghi2018icassp}. 
The objective in the latter methodology is to preserve \textit{semantic} or \textit{distance similarity} between the original space and transformed space. 
Subsequent research showed that quantization-based solutions are preferred in terms of query time, space cost and search accuracy \cite{wang2017survey}.

\begin{figure}
\centering
\includegraphics[scale=0.75]{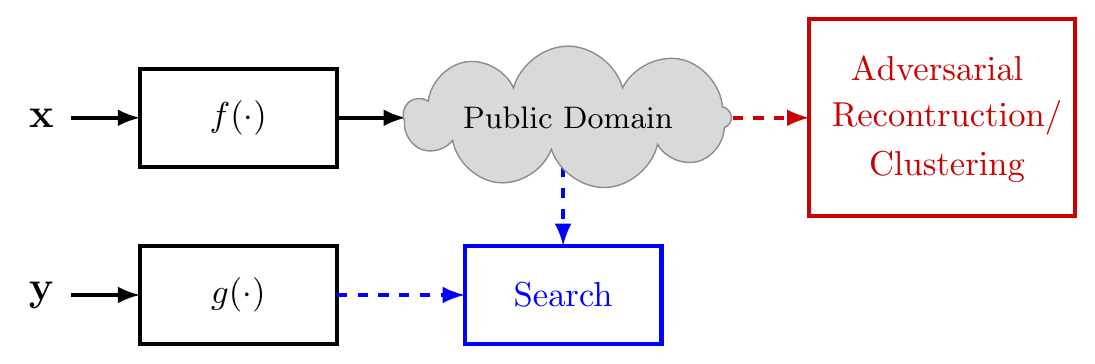}
\caption{The general block diagram of our framework.}
\label{Fig:GeneralDiagram}
\end{figure}

%

Concretely, this work brings the following contributions:
(1)
We consider the methodology of learning to hash for privacy-preserving proximity search which entails minimum information loss for authorized users. 
The authorized parties can \textit{purify} the ambiguation noise using the \textit{shared secrecy} based on the support intersection of sparse codes \cite{SohrabBehrooz2020icassp}.  
\\
(2) 
We adopt a notion of fairness addressed in \cite{har2019near}, but in a privacy-preserving setup.  
Our notion of fairness differs from machine learning algorithms where the goal is to handle the bias introduced at the training phase. 
We consider the bias in the \textit{stored data} and \textit{querying response}.
By doing this, any point in the neigborhood has an equal chance to be chosen. 
Moreover, in some cases, 
it may suffice to return any of the points in the near neighborhood, rather than the computationally expensive nearest one.
The equiprobable nearby scheme can also be utilized in privacy protection mechanisms.  That is, instead of reporting the nearest-neighbor, which leaks more information, the service provider just sends back a random or a typical data point close to the query point.  

In comparison to \cite{boufounos2011secure, weng2016privacy},  our work has the following fundamental differences: 
%
a) 
In \cite{boufounos2011secure, weng2016privacy, kenthapadi2012privacy, rane2013privacy}, they utilized a dimensionality reduction transform with random entries, while our sparsifying transform may keep, extend, or reduce the dimension of the original data. 
Moreover, our transform is learned using the sparsifying transform problem to ensure an optimal sparse representational that is information preserving in general, whereas the transform in \cite{boufounos2011secure, weng2016privacy, kenthapadi2012privacy} might preserve the distances only under certain conditions of the Johnson-Lindenstrauss Lemma. 
b)  
In \cite{boufounos2011secure, weng2016privacy, rane2013privacy}, the codes are dense and binary, whereas in our method the codes are sparse (and possibly ternary), which form a basis of our ambiguation framework. 
Last but not the least, the embedding based on universal quantization scheme \cite{boufounos2011secure} has information leakage in terms of clustering, i.e., the curious server still can perform clustering on data points.  
%
Moreover, we impose no restrictions on the input data, i.e., we assume that as an input we might have raw data, extracted features using any known hand crafted methods, aggregated local descriptors based on BoW, FV, VLAD \cite{jegou2009burstiness, perronnin2007fisher, jegou2010aggregating}, etc., or from the last layers of deep nets \cite{babenko2014neural}, or the latent space of auto-encoders \cite{kingma2014auto}. 
We apply our model on the latent representation of a designed network in~\cite{SohrabBehrooz2020icassp}. 
%
%



Throughout this paper, 
superscript $(\cdot)^T$ stands for the transpose. 
Vectors and matrices are denoted by boldface lower-case ($\mathbf{x}$) and upper-case ($\mathbf{X}$) letters, respectively. 
We consider the same notation for a random vector $\mathbf{x}$ and its realization. The difference should be clear from the context. 
$x_i$ denotes the $i$-th entry of vector $\mathbf{x}$. 
For a matrix $\mathbf{X}$, 
$\mathbf{x}{\left(  j \right)}$ denotes the $j$-th column of $\mathbf{X}$. 
We use the notation $\left[ N \right]$ for the set $\{ 1, 2, ..., N\}$.

  



\section{Preliminaries}
\label{Sec:Preliminaries}
 


\subsection{Problem Setup}


%
Consider a three-party data release scenario involving (a) a data owner, (b) data users, and (c) a service provider (server). 
The data owner possesses database $\mathbf{X} \! = \! \left[ \mathbf{x}(1), \cdots, \mathbf{x}(M) \right]$ consisting of $M$ data points $\mathbf{x}(m) \! \in \! \mathbb{R}^N$, $m \in \left[ M \right]$. 
The database is used to offer some utility 
for the \textit{authorized} data users. 
%
%
The data users seek some utility from the data owner based on their query 
$\mathbf{y}$. 
The server provides a pre-determined service to the data users on behalf of the data owner. 
We assume that both the server and data users are honest-but-curious, which we consider them as an adversary. 
The service provider may try to infer some information about the original data collection $\mathcal{X}$ from the disclosed public storage and/or the querying data sent to the server. For instance, the server may estimate the original data from the disclosed representations and query, or may establish links between the closet entries in database. 
The data users may try to infer some information about the public representations and/or the original data via multiple varied queries to guess the data manifold by inspecting the returned responses. A general diagram of our framework is depicted  in Fig.~\ref{Fig:GeneralDiagram}.  

Therefore, we study the problem of disclosing database $\mathbf{X}$ to a third-party (public storage) in order to drive some utility, in terms of near neighbor search, for the authorized data users based on the public representations while, at the same time, protect the privacy of the data owner (against the honest-but-curious server and data users) and data users (against the honest-but-curious server).




\subsection{Fair Near Neighbor}


Let $\left( \mathcal{S}, d_{\mathcal{S}} \right)$ be a metric space and let $\mathcal{X} \subseteq \mathcal{S}$ be a set of $M$ data points. 
%
%
Let $B_{\mathcal{S}}\left( \mathbf{c}, r \right) = \{ \mathbf{x} \in \mathcal{S} \mid d_{\mathcal{S}} \left(\mathbf{c}, \mathbf{x} \right) \leq r  \}$ be the closed ball of radius $r > 0$ around a point $\mathbf{c} \in \mathcal{S}$. 
Let $N\left( \mathbf{c},r \right) = B_{\mathcal{S}} \left( \mathbf{c}, r \right) \cap \mathcal{X}$ be the $r$-neighborhood of $\mathbf{c}$ in $\mathcal{X}$, with the size $\vert N\left( \mathbf{c},r \right) \vert $. 


\theoremstyle{definition}
\begin{definition}{Fair Near Neighbor (FNN) \cite{har2019near}.}
Given a data set $\mathcal{X} \subseteq \mathcal{S}$ of $M$ data points, a parameter $r > 0$, and query point $\mathbf{y}$, the goal is to find a data point $\mathbf{x} \in N\left( \mathbf{y}, r \right) $ with probability $\mu$, where 
$1 / \! \left( \vert N \left( \mathbf{y}, r\right)\vert ( 1 + \epsilon) \right)  \! \leq \!  \mu \!  \leq \! (1 \! + \epsilon)/ \vert N\left( \mathbf{y}, r \right) \! \vert $, 
i.e., $\mu$ is an approximately uniform probability distribution.
\end{definition}

\begin{figure}[!t]
\centering
\includegraphics[scale=0.4]{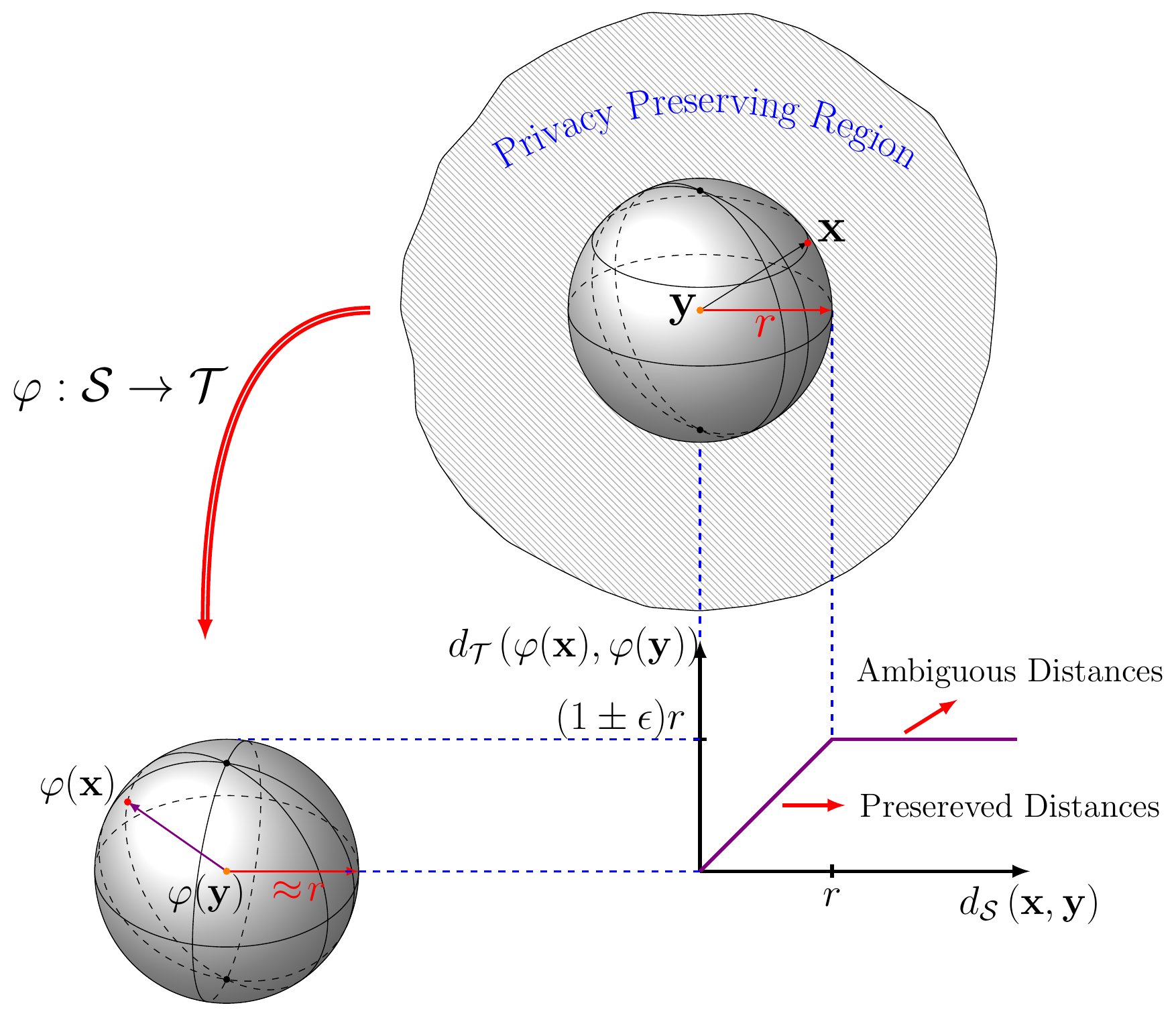}
\vspace{-3pt} 
\caption{Visualization of desired property of mapping scheme in privacy-preserving near neighbor search setup.}
\label{Fig:DesiredMappingScheme}
\vspace{-5pt} 
\end{figure}



\subsection{Fair Privacy-Preserved Approximate Near Neigbor}


Let $\left( \mathcal{T}, d_{\mathcal{T}} \right)$ be a metric space where $d_{\mathcal{T}} \left( \mathbf{a}, \mathbf{b} \right), \forall \mathbf{a}, \mathbf{b} \in \mathcal{T}$ is defined on $\mathrm{supp} (\mathbf{a}) = \{ l \in [L] : a_l \neq 0 \}$. Let $\mathcal{P} \subseteq \mathcal{T}$ be a set of $M$ data points. 
Let $B_{\mathcal{T}}\left( g(\mathbf{y}), r \right) = \{ f(\mathbf{x}) \in \mathcal{T} \mid d_{\mathcal{T}} \left( g(\mathbf{y}), f (\mathbf{x}) \right) \leq r \}$. 
Now, we define the \textit{Fair Privacy-preserved Approximate Near Neighbor}  method as follows:
%
%
%


\theoremstyle{definition}
\begin{definition}{Fair Privacy-preserved Approximate Near-Neighbor (FPANN).}
Given a data set $\mathcal{X} \subseteq \mathcal{S}$ of $M$ data points, a parameter $r > 0$, 
the goal is to design a randomized privacy-preserving data release mechanism $f : \mathcal{X} \rightarrow \mathcal{P}$ and 
(randomized) query processing $g: \mathcal{S} \rightarrow \mathcal{T}$
such that for a given \textit{authorized} query $ \mathbf{y}$ one can report 
a point $f \left( \mathbf{x} \right) \in N_{\mathcal{T}} \left( g(\mathbf{y}), r \right)$ with probability $\mu_p$, where $N_{\mathcal{T}}\left( g(\mathbf{y}), r \right) = B_{\mathcal{T}}\left( g(\mathbf{y}), r \right) \cap \mathcal{P}$ be the approximate $r$-neighbor of $g(\mathbf{y})$ in $\mathcal{P}$, and 
$1 / \left( \vert N\left( g(\mathbf{y}), r \right) \! \vert \, (1 + \epsilon) \right) \leq  \mu_p  \leq (1 + \epsilon)/ \vert N\left( g(\mathbf{y}), r \right) \! \vert$. 
\end{definition}



\begin{definition}{$\left( \beta, \gamma \right)$-recoverable privacy mechanism.}
 For $0 \leq \gamma \leq 1$ and given authorized query $\mathbf{y}^{\mathrm{auth}}$, unauthorized query $\mathbf{y}^{\mathrm{unauth}}$ and $\beta > 0$, a privacy-preserving data release mechanism $f: \mathcal{X} \rightarrow \mathcal{P}$ is $\left( \beta, \gamma \right)$-recoverable if:\vspace{-1pt}
\begin{eqnarray}
(i):   \quad P_e^{\mathrm{auth}}  \!\! &=& \!\! \mathrm{Pr} \left[ \mathbb{E} \left[ d \left( \mathbf{x}, \mathbf{\hat{x}} \right) \right] \leq \beta \mid g \left( \mathbf{y}^{\mathrm{auth}} \right) \right] < \gamma ,\nonumber \\
(ii): P_e^{\mathrm{unauth}}   \!\! &=&  \!\! \mathrm{Pr} \left[ \mathbb{E} \left[ d \left( \mathbf{x}, \mathbf{\hat{x}} \right) \right] \leq \beta \mid g \left( \mathbf{y}^{\mathrm{unauth}} \right) \right]\geq  \gamma ,\nonumber 
\end{eqnarray}%
where $g (\cdot)$ is the data user's query function to service provider. 
\end{definition}



\section{Proposed Framework}
\label{Sec:ProposedFramework}
 
\vspace{-2pt}

 
\subsection{Framework Overview}

 %
Our framework is composed of the following steps:

\noindent
%
1) \textit{Preparation at Owner Side:}
The owner generates the sparse codewords from the data that s/he owns using the  
\textit{trained sparsifying transform}. 
Next, he shares the privacy-protected sparse codebook with the service provider (server). 
Following Kerckchoffs's Principle in cryptography, the data owner makes the learned sparsifying transform publicly available. 
%
 
\noindent
2) \textit{Indexing at Server Side:}
The server indexes the received sparse codes in a database. 
%
 
\noindent
3) \textit{Querying at Data User Side:}
The data user generates a sparse representation from his query data using the  shared transform. 
Then, the client sends a function of his sparse representation to the server. 
%
 
\noindent
4) \textit{Near Neighbor Search at Server Side:} 
Given the requested probe, 
the server runs a near neighbor search to find the stored sparse codes that are most similar (close) to the probe. 
Finally, based on the pre-determined service to the data users, the server sends back an answer to the data user. 

Next, we describe in more detail the fundamental elements of our mechanism.



\subsection{Sparse Data Representation}%
\label{sec:SparseDataRepresentation-FirstKeyElement}


%
%

The goal of sparsification is to obtain an information-preserving sparse representation of the original data. 
Our sparsifying transform consists of a linear mapper followed by an element-wise nonlinearity. We consider a joint learning problem to obtain the sparsifying transform $\mathbf{W} \in \mathbb{R}^{L \times N}$ as well as the sparse codebook $\mathbf{A} \in \mathbb{R}^{L \times M}$ that can be formulated as:\vspace{-1pt}
\begin{equation}\label{Eq:JointLearning}
\!\!\! \big(   \mathbf{\hat{W}}, \! \mathbf{\hat{A}}  \big) \! = \! \mathrm{arg} \!  \mathop{\min}_{\left( \mathbf{W}, \mathbf{A} \right)} \!  {\| \mathbf{W} \mathbf{X} \!  - \!  \mathbf{A} \|}_F^2 \! + \! \beta_1 \Omega_1 \!  \left( \mathbf{W} \right) \! + \!  \beta_2 \Omega_2 \! \left( \mathbf{A} \right),
\end{equation}
where $\beta_1 \geq 0$ and $\beta_2 \geq 0$ are regularization parameters, $\Omega_1 \!  \left( \mathbf{W} \right)= (\frac{1}{\beta_{1,1}}\Vert {\bf W} \Vert_{F}^2+\frac{1}{\beta_{1,2}}\Vert {\bf W}{\bf W}^T- {\bf I} \Vert_F^2-\frac{1}{\beta_{1,3}}\log \vert \det {\bf W}^T{\bf W} \vert )$ penalizes the information loss
in order to avoid trivial solutions, 
and $\Omega_2 \! \left( \mathbf{A} \right)$ is the sparsity constraint on the compressed codebook $\mathbf{A}$ \cite{ravishankar2013learning}. 
The term ${\| \mathbf{W} \mathbf{X} \!  - \!  \mathbf{A} \|}_F^2$ is a sparsification error, which represents the deviation of the transformed data from the exact sparse representation in the transformed domain \footnote{We  refer  the  reader  to \cite{gheisari2020joint, gheisari2019group, gheisari2019aggregation} for applications in group membership verification.}
Our algorithm for solving \eqref{Eq:JointLearning} alternates between a $\ell_0$-``norm"-based \textit{sparse coding step}, and a non-convex \textit{transform update step} \cite{Kostadinov2018:EUVIP}. 
Therefore, one can write the closed-form formulation of the \textit{encoder} as \cite{Kostadinov2018:EUVIP, Razeghi2017wifs}:
\begin{equation}\label{Eq:Encoder}
\mathbf{a}(m) = \varphi \left( \mathbf{x}(m) \right) = \psi_{\lambda} \left( \mathbf{W}\mathbf{x}(m)\right), \forall m \in \left[ M \right], 
\end{equation}
where $\psi_{\lambda} \! \left( \mathbf{f} \right) \! = \! \mathds{1}_{\vert f_l  \vert \ge \lambda } {\mathbf{f}}, \forall l \! \in \! \left[ L\right], \lambda \geq 0$ and $\mathbf{a} (m)$ is $S_x$-sparse, i.e., $\Vert \mathbf{a} (m) \Vert \!  \approx \! S_x, \forall m \in \left[ M\right]$.  
%
%
%
The decoder (reconstruct mapper) $\mathbf{R} \in \mathbb{R}^{N \times L}$ can be formulated as follows:\vspace{-2pt}
\begin{eqnarray}\label{ReconstructionRegulization}
\!\!\!\!\!\!\!\!\!\!\!\!  \mathop{\min}_{\mathbf{R}} \! \! \! \! &&\! \! \! \!   {\|  \mathbf{R} \mathbf{A} \!  - \! \mathbf{X}   \|}_F^2 + \! \beta_R { \| \mathbf{R} \! - \!  {\left( \mathbf{W}^T \mathbf{W} \!  +\!  \beta \mathbf{I} \right)}^{\! -1}  \mathbf{W}^T\!  \| }_F^2, \nonumber \\
 \mathrm{st:} \! \! \! &&\!\!\!\!   \mathbf{R}^T \mathbf{R} = \mathbf{I},
\end{eqnarray}
where $\mathbf{A} \in \mathbb{R}^{L \times M}$ is sparse codebook, $\mathbf{X} \in \mathbb{R}^{M \times N}$ is original data points and $\mathbf{W} \in \mathbb{R}^{L \times N}$ is encoder transform. Since $\mathbf{R}$ has orthonormal columns, we have 
$ {\|  \mathbf{R}\! \mathbf{A} \! - \! \mathbf{X}   \|}_F^2 \! \! \!  = \! \!\! \! \! \!  \mathrm{tr}\! \left[ \mathbf{X}^T \mathbf{X} \! -\! 2 \mathbf{X}^T \mathbf{R} \mathbf{A} \! + \! \mathbf{A}^T \! \mathbf{A} \right]\!, 
 { \| \mathbf{R} \! -\! \!  {\left( \mathbf{W}^T \mathbf{W} \! \! +\!  \beta \mathbf{I} \right)\!}^{\! -1}  \mathbf{W}^{  T}  \| }_F^2  \! \! \! \!   =  \! \!  \! \! \mathrm{tr} \! \left[  \mathbf{I} - 2 \mathbf{C}^T \mathbf{R} + \mathbf{C}^T \mathbf{C} \right]$, 
where $\mathbf{C} =  {\left( \mathbf{W}^T \mathbf{W} + \beta \mathbf{I} \right)}^{-1} \mathbf{W}^T$. Consequently \eqref{ReconstructionRegulization} is equivalent to the problem of maximizing $\mathrm{tr}\left[ \mathbf{X}^T \mathbf{R} \mathbf{A} \right] + \beta_R \mathrm{tr}\left[ \mathbf{C}^T \mathbf{R}  \right] = \mathrm{tr}\left[ \left( \mathbf{A} \mathbf{X}^T + \beta_R \mathbf{C}^T \right) \mathbf{R}  \right]$. Considering the Singular Value Decomposition $\mathbf{A} \mathbf{X}^T + \beta_R \mathbf{C}^T = \mathbf{U} \mathbf{\Sigma} \mathbf{V}^T$, this formulation reduces to 
%
$\mathrm{tr} \left[ \mathbf{U} \mathbf{\Sigma} \mathbf{V}^T \mathbf{R} \right] = 
\mathrm{tr} \left[ \mathbf{\Sigma} \mathbf{Z} \right] 
= \sum_{i} z_{ii} \Sigma_{ii} \leq \sum_{i} \Sigma_{ii},$ 
where $\mathbf{Z} = \mathbf{V}^T \mathbf{R} \mathbf{U}$. Note that the last inequality holds because $\mathbf{Z}$ is an orthonormal matrix, and $\sum_{j} z_{ij}^2 = 1$, $z_{ii} \leq 1$. Therefore, the maximum can be achieved if $\mathbf{Z} = \mathbf{I}$, i.e., closed form solution is $\mathbf{R} = \mathbf{U} \mathbf{V}^T$, where $\mathbf{A} \mathbf{X}^T + \beta_R \big( { \left( \mathbf{W}^T \mathbf{W} + \beta \mathbf{I} \right) }^{-1} \mathbf{W}^T \big) ^T = \mathbf{U} \mathbf{\Sigma} \mathbf{V}^T$.



\subsection{Ambiguation Mechanism}%
\label{sec:PrivacyAmplification}
%
%
The idea of ambiguation is to add (pseudo) random noise to the orthogonal complement, i.e., non-informative components of the sparse code. 
The integration of `sparse lossy coding' with `ambiguation' introduces a generalized randomization technique, namely Sparse Coding with Ambiguation (SCA) \cite{Razeghi2017wifs}. 
The SCA provides an information-theoretically and computationally private mechanism. 
The information-theoretical privacy guarantee originate from the lossy compression induced at the sparsification stage, and the computational privacy guarantee originate from ambiguation stage. The curious server faces a combinatorial complexity budget requirement to guess the informative components. 
The ambiguation noise is required to have the same distribution 
as the sparse codes, to guarantee being indistinguishable from its statistical properties. 
We refer the reader to \cite{Razeghi2017wifs} for more details. 
The randomized privacy-preserving data release mechanism $f: \mathcal{X} \rightarrow \mathcal{P} \subseteq \mathcal{T}$ can be formulated as:\vspace{-2pt}
\begin{equation}
\mathbf{p} (m) = f \! \left( \mathbf{x}(m) \right) =  \varphi \! \left( \mathbf{x}(m) \right) \oplus \mathbf{n}_{\mathrm{supp}}^p 
, \;    \forall m \in \left[ M\right], 
\end{equation}
where ${\Vert \mathbf{n}_{\mathrm{supp}}^p \Vert}_0 \approx S_p$.

Given a query point $\mathbf{y}$, the (randomized) query release mechanism $g: \mathcal{S} \rightarrow \mathcal{T}$ can be formulated as:\vspace{-2pt}
\begin{equation}
\mathbf{q}  = g \! \left( \mathbf{y} \right) = \varphi \left( \mathbf{y} \right) \oplus \mathbf{n}_{\mathrm{supp}}^q,
\end{equation}
where ${\Vert \mathbf{n}_{\mathrm{supp}}^q \Vert}_0 \leq  S_q$, $0 \leq S_q \leq S_p$. If $S_q = 0$, the query is disclosed as in-the-clear sparse code without ambiguation noise. 
Let us consider two hypotheses for near neighbor search as follows. 
$\mathcal{H}_1$: The authorized query is related to one of the $M$ data points in the database. For instance, it is a noisy version of one data point. 
$\mathcal{H}_0$: The unauthorized query is not related to any data point. For instance, it is synthetic query generated by an adversary.



 
\subsection{Near Neighbor Search}%
\label{sec:NearNeighborSearch}
%
%
%
%
The near neighbor search is performed in latent space $\mathcal{T}$. 
Given a data set 
$\mathcal{P}$ of $M$ embedded disclosed representations $\{ \mathbf{p} (m) \}, m \in \left[M \right]$, a parameter $r$, and embedded query point $\mathbf{q}$, the service provider performs approximate near neighbor search and report a point randomly and uniformly from $B_{\mathcal{T}} \left( \mathbf{q}, r \right) \cap \mathcal{P}$.



\section{Discussion}
\label{Sec:PrivacyUtilityGuarantees}

\vspace{-2pt}

We now discuss various properties of our method. 
One desired property of an embedding scheme in a privacy-preserving near neighbor search is to preserve distance information only up to a specified radius, while quickly flattening after this distance threshold. Therefore, from one hand, the information rate is spent in encoding local distances, and from the other hand, the curious server/data user cannot recover any distance information about signals that are far apart. 
Fig.~\ref{Fig:DesiredMappingScheme} visualizes this local isometric mapping, where $d_{\mathcal{S}} \left( \cdot, \cdot \right)$ and $d_{\mathcal{T}} \left( \cdot, \cdot \right)$ denote the distance measure in original domain and transform domain, respectively. 

\begin{figure*}[!t]
    \centering
     \begin{subfigure}[h]{0.258\textwidth}
        \includegraphics[width=\linewidth, height=3.4cm]{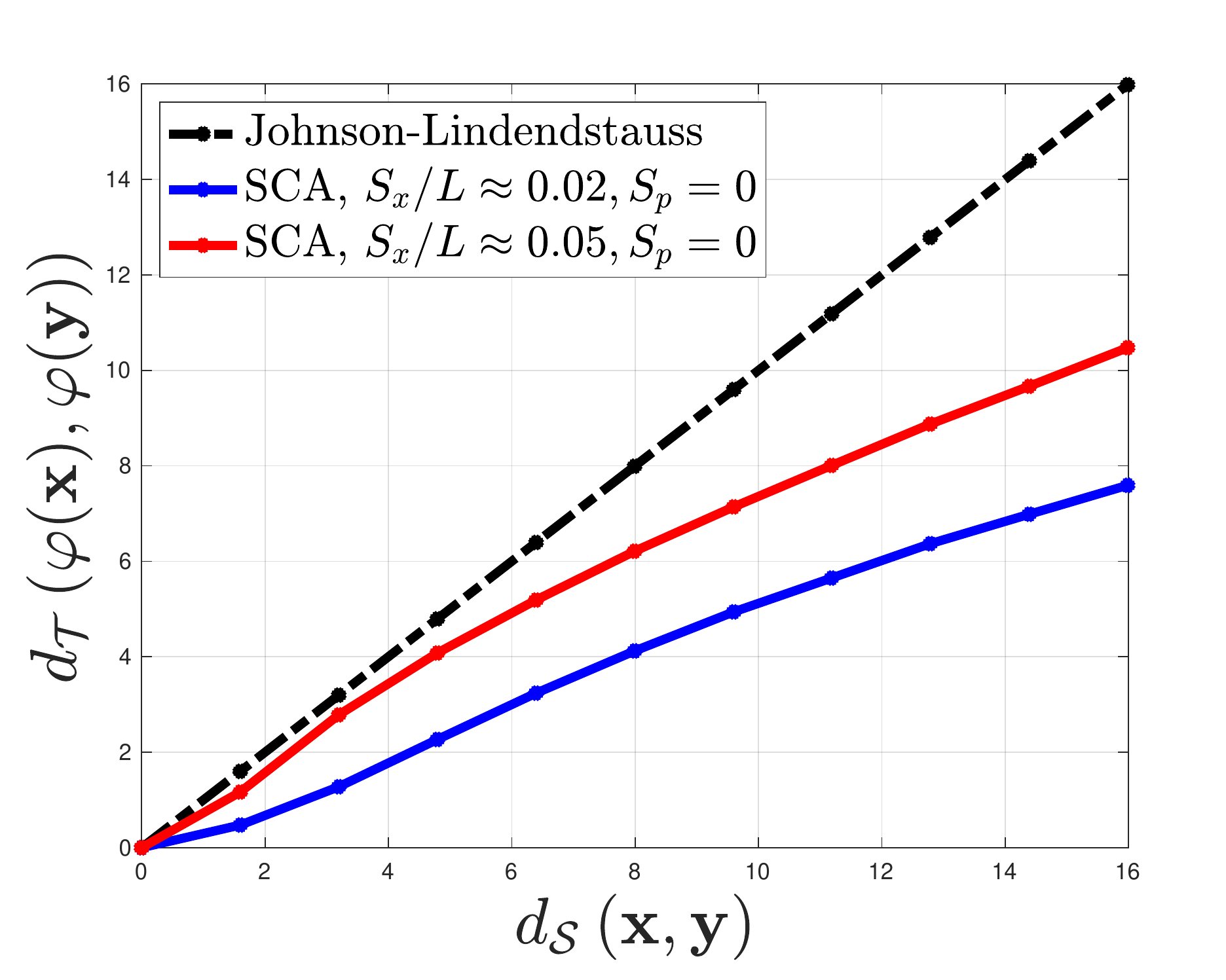}%
        \vspace{-5pt}
         \caption{ }
          \label{fig:DiatncePresrving}
     \end{subfigure}
     ~
     \begin{subfigure}[h]{0.258\textwidth}
        \includegraphics[width=\linewidth, height=3.4cm]{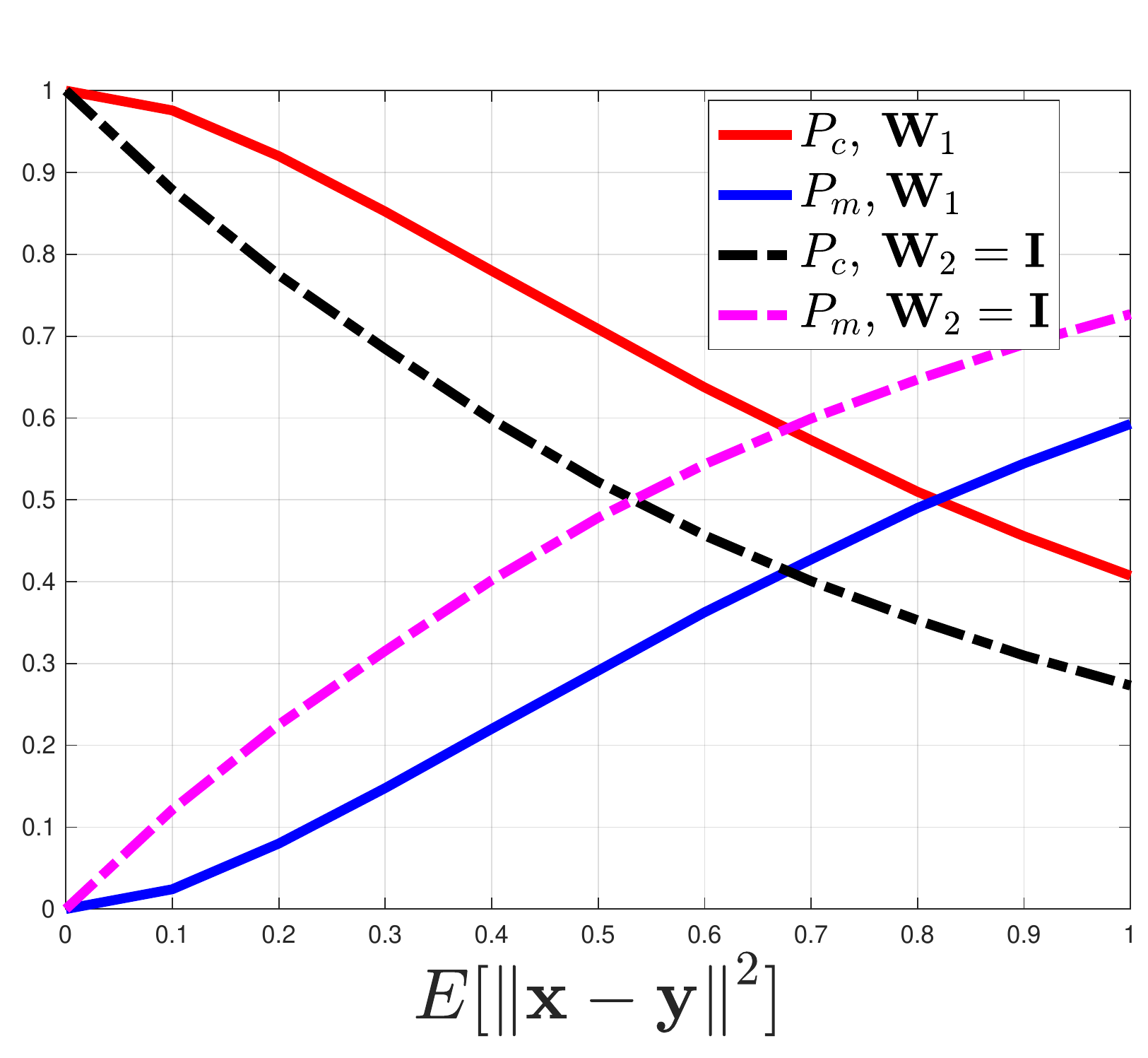}%
        \vspace{-5pt}
         \caption{ }
          \label{fig:PcorrectPmiss}
     \end{subfigure}
     ~
     \begin{subfigure}[h]{0.258\textwidth}
        \includegraphics[width=1.08\linewidth, height=3.4cm]{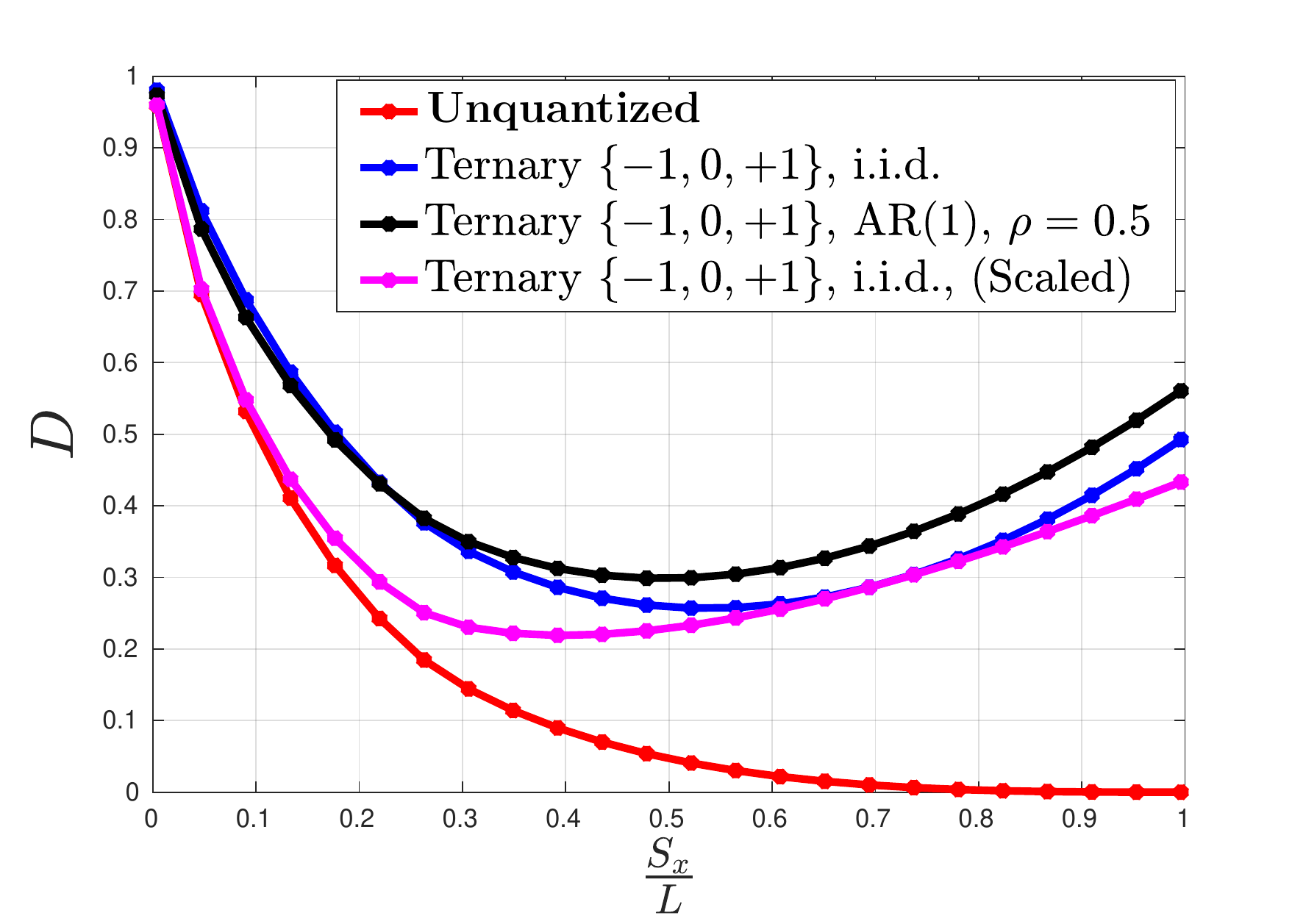}%
        \vspace{-5pt}
         \caption{ }
          \label{fig:DistortionSparsityAuthorized}
     \end{subfigure}
     
     \begin{subfigure}[h]{0.258\textwidth}
        \includegraphics[width=\linewidth, height=3.4cm]{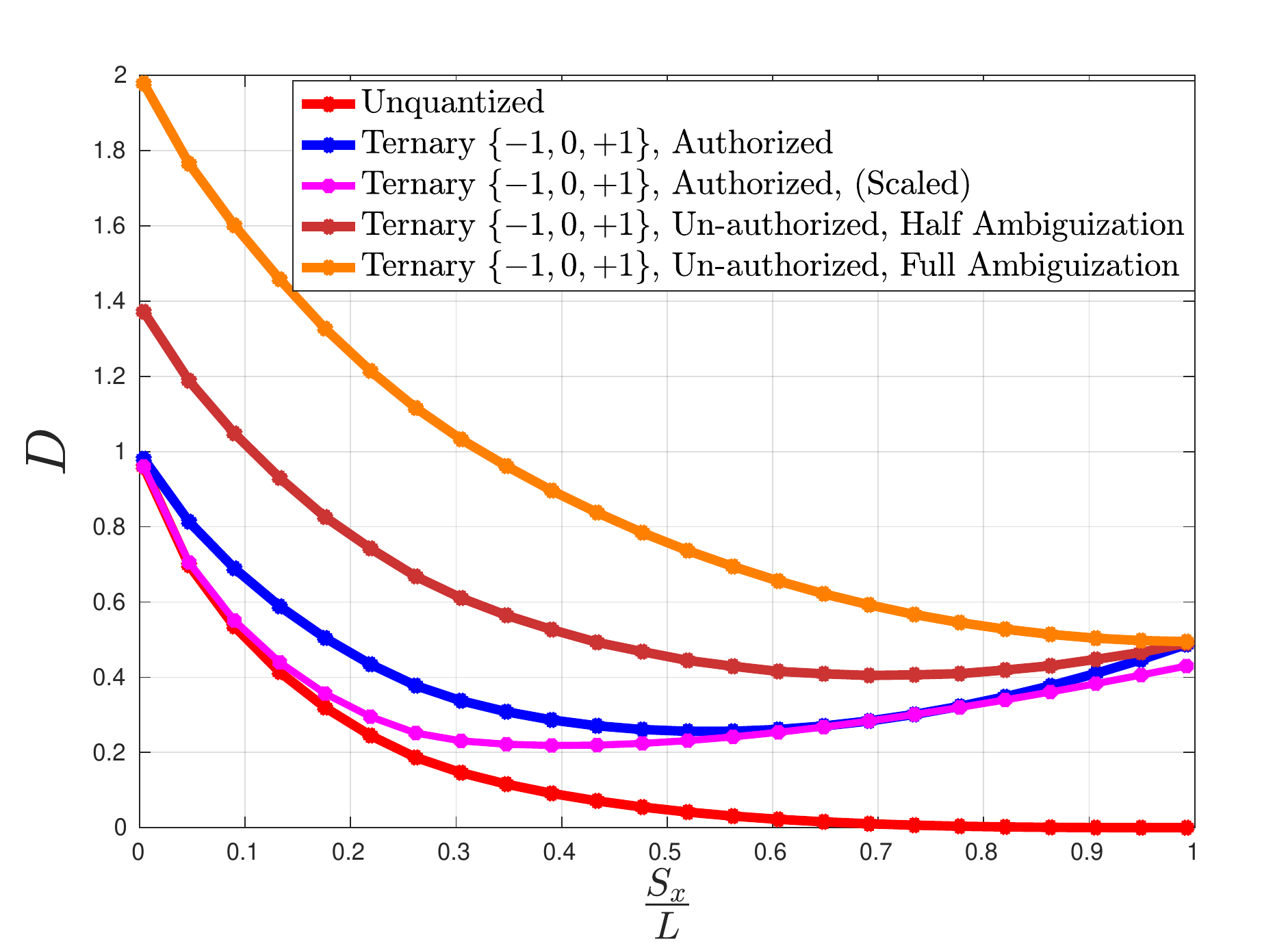}%
        \vspace{-5pt}
         \caption{ }
          \label{fig:DistortionSparsityUnauthorized}
     \end{subfigure}
     ~
     \begin{subfigure}[h]{0.258\textwidth}
        \includegraphics[width=\linewidth,height=3.3cm]{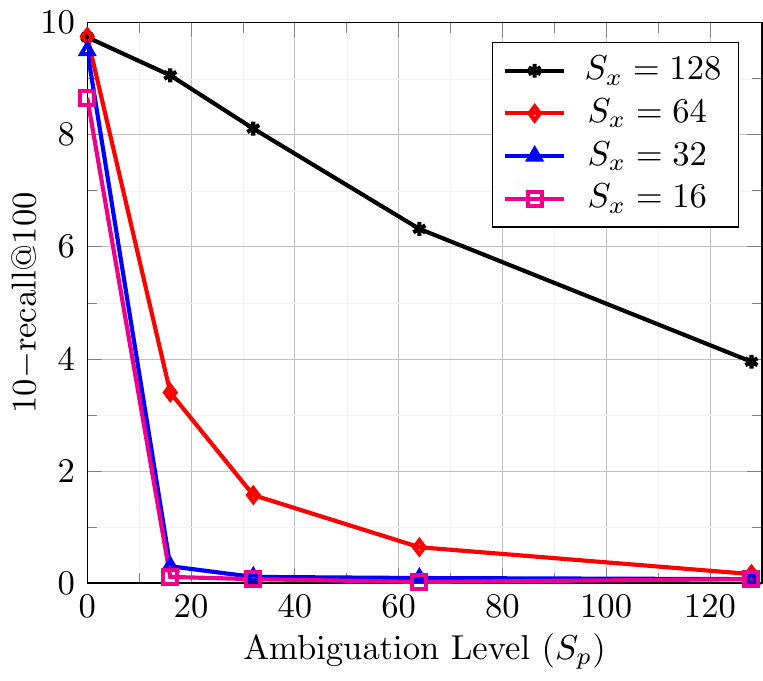}%
        \vspace{-8pt}
         \caption{ }
          \label{fig:10_recall_100}
     \end{subfigure}   
     ~
     \begin{subfigure}[h]{0.258\textwidth}
        \includegraphics[width=\linewidth,height=3.3cm]{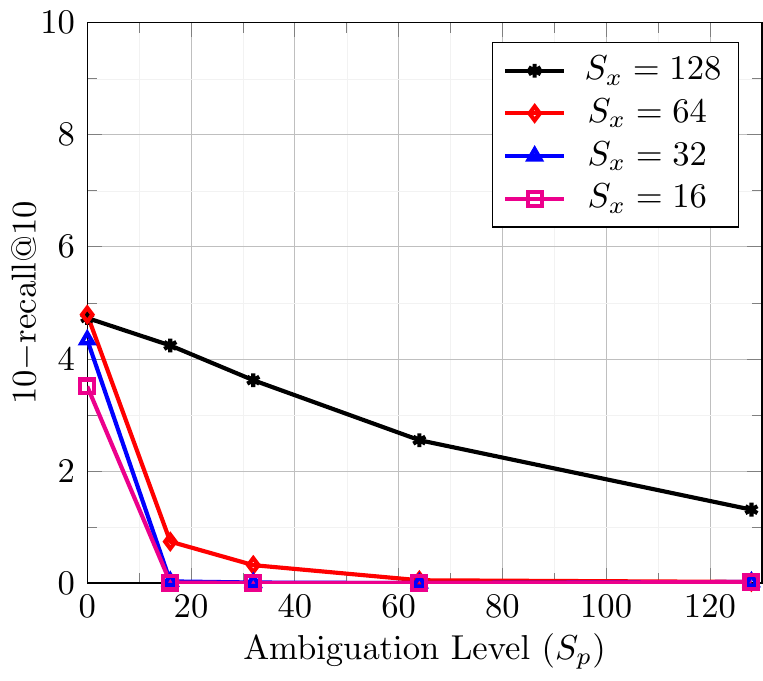}%
        \vspace{-8pt}
         \caption{ }
          \label{fig:10_recall_10}
     \end{subfigure}  
     \caption{a) local distance preserving; b) local robustness; (c) and (d): Comparison of distortion-sparsity behavior for c) authorized and d) unauthorized parties; (e) and (f) $R-$recall$@T$ curves for a subset of 10K CelebA images of $3 \times 128 \times 128$.}
     \label{Fig:performance}
\end{figure*}

Suppose $\mathbf{x} \! \sim \! \mathcal{N} \! \left( \boldsymbol{0}, \sigma_{\mathbf{x}} \mathbf{I}_N  \right)$ and $\mathbf{y}^{\mathrm{auth}} \! =\! \mathbf{x} + \mathbf{z}$, where $\mathbf{z} \sim \mathcal{N} \! \left( \boldsymbol{0}, \sigma_{\mathbf{z}} \mathbf{I}_N  \right)$, and where $\mathbf{x} \in \mathbf{X}$. 
Fig.\ref{fig:DiatncePresrving} depicts the behaviour of our embedding for two sparsity levels and compare them with linear embeddings which preserve all distances equally. 
Let us define:\vspace{-3pt}
\begin{eqnarray}
P_c \! &=& \! \frac{1}{M. S_x} \sum \mathrm{Pr} \{   \mathrm{supp}\left( \varphi (\mathbf{x}) \right) = \mathrm{supp}\left( \varphi (\mathbf{y}) \right)\}, \nonumber \\
P_m \! &=& \! \frac{1}{M. S_x} \sum  \mathrm{Pr} \{   \mathrm{supp}\left( \varphi (\mathbf{x}) \right) \neq \mathrm{supp}\left( \varphi (\mathbf{y}) \right)\}. \nonumber 
\end{eqnarray}
Fig.\ref{fig:PcorrectPmiss} illustrates the probability of correct support and missed support for the learned linear map $\mathbf{W}_1$ (problem \eqref{Eq:JointLearning}) and $\mathbf{W}_2 = \mathbf{I}$. The learned transform outperforms in local distances.

\begin{table*}[b!]
\centering
\resizebox{0.85\linewidth}{!}{%
\begin{tabular}{|cc|ccccc||ccccc|}
\cline{3-12}
\multicolumn{1}{c}{} & & \multicolumn{5}{c||}{$S_p = 0$} &  \multicolumn{5}{c|}{$S_x = 16$}\\
\cline{3-12}
\multicolumn{1}{c}{} & & $S_x = 1$ &  $S_x = 2$ &  $S_x = 4$ &  $S_x = 8$ & $S_x = 16$ & $S_p = 0$ &  $S_p = 8$ &  $S_p = 16$ &  $S_p = 24$ & $S_p = 32$ \\  \cline{1-2}
 \multirow{2}{*}{MNIST} &
 \multicolumn{1}{|c|}{MSE} &  \cellcolor{gray!45}{$0.7331$} &  \cellcolor{gray!37}{$0.6351$} & \cellcolor{gray!29}{$0.5101$} &  \cellcolor{gray!21}{$0.3758$} &  \cellcolor{gray!14}{$0.2726$} & %
 \cellcolor{gray!14}{$0.2726$} &  \cellcolor{gray!21}{$0.4030$} & \cellcolor{gray!29}{$0.4963$} &  \cellcolor{gray!37}{$0.5607$} &  \cellcolor{gray!45}{$0.6120$} \\ \cline{2-2}
& \multicolumn{1}{|c|}{SSIM} &  \cellcolor{gray!14}{ $0.5838$} & \cellcolor{gray!21}{ $0.6712$} & \cellcolor{gray!29}{$0.7694$} & \cellcolor{gray!37}{ $0.8628$} & \cellcolor{gray!45}{ $0.9335$} & %
\cellcolor{gray!45}{$0.9335$} & \cellcolor{gray!37}{$0.8124$} & \cellcolor{gray!29}{$0.7330$} & \cellcolor{gray!21}{$0.6750$} & \cellcolor{gray!14}{$0.6290$} \\   \hline
\multirow{2}{*}{F-MNIST} &
\multicolumn{1}{|c|}{MSE}  &  \cellcolor{gray!45}{$0.5355$} & \cellcolor{gray!37}{$0.4298$} & \cellcolor{gray!29}{$0.3368$} & \cellcolor{gray!21}{$0.2657$} & \cellcolor{gray!14}{$0.2236$} & %
\cellcolor{gray!14}{$0.2236$} & \cellcolor{gray!21}{$0.2775$} & \cellcolor{gray!29}{$0.3255$} & \cellcolor{gray!37}{$0.3635$} & \cellcolor{gray!45}{$0.4026$}\\ \cline{2-2}
& \multicolumn{1}{|c|}{SSIM} &  \cellcolor{gray!14}{$0.4926$} & \cellcolor{gray!21}{$0.5989$} & \cellcolor{gray!29}{$0.6976$} & \cellcolor{gray!37}{$0.7753$} & \cellcolor{gray!45}{$0.8242$} & %
\cellcolor{gray!45}{$0.8242$} & \cellcolor{gray!37}{$0.7506$} & \cellcolor{gray!29}{$0.6835$} & \cellcolor{gray!21}{$0.6351$} & \cellcolor{gray!14}{$0.5914$} \\   \hline
\multirow{2}{*}{CIFAR-10} &
\multicolumn{1}{|c|}{MSE} & \cellcolor{gray!45}{$0.3993$}  & \cellcolor{gray!37}{$0.3439$} & \cellcolor{gray!29}{$0.2741$} & \cellcolor{gray!21}{$0.2061$} & \cellcolor{gray!14}{$0.1593$} & %
 \cellcolor{gray!14}{$0.1593$}  & \cellcolor{gray!21}{$0.2066$} & \cellcolor{gray!29}{$0.2420$} & \cellcolor{gray!37}{$0.2703$} & \cellcolor{gray!45}{$0.2920$} \\ \cline{2-2}
& \multicolumn{1}{|c|}{SSIM} & \cellcolor{gray!14}{$0.4711$}  & \cellcolor{gray!21}{$0.5429$} & \cellcolor{gray!29}{$0.6342$} & \cellcolor{gray!37}{$0.7289$} & \cellcolor{gray!45}{$0.8002$} & %
\cellcolor{gray!45}{$0.8002$}  & \cellcolor{gray!37}{$0.7188$} & \cellcolor{gray!29}{$0.6619$} & \cellcolor{gray!21}{$0.6179$} & \cellcolor{gray!14}{$0.5883$}\\ \hline
\end{tabular}
} 
\vspace{-4pt}
\caption{Reconstruction quality vs sparsity and ambiguation levels. 
}
\label{table:RD}
\end{table*}


\subsection{Reconstruction Leakage}


A potential threat is that the adversary may try to reconstruct the original data points  from the disclosed representations. 
To get insight into the SCA model, we firstly provide the results on a synthetic database establish its connection to classical Shannon rate-distortion theory. 
Next, we validate our model on real image databases. 
For the sake of completeness, we also bring the results provided in \cite{Razeghi2018icassp} on synthetic i.i.d data. 
Note that the sparsity level $S_x$ controls the information encoding rate, or equivalently, the distinguishability of data points in the transform domain. 
The ambiguation level $S_p$ controls the imposed randomness to the informative data.

Fig.~\ref{fig:DistortionSparsityAuthorized} and Fig.~\ref{fig:DistortionSparsityUnauthorized} illustrate and compare distortion-sparsity behavior at \textit{authorized} and \textit{unauthorized} parties, respectively. 
Fig.\ref{fig:DistortionSparsityAuthorized} depicts reconstruction fidelity for four cases: 1) unquantized sparsifying encoding \eqref{Eq:Encoder}, 2) Sparse Ternary Coding (STC) for independent and identically distributed (i.i.d.) data \cite{Razeghi2017wifs}, 3) STC for i.i.d. data which re-scaled in original domain \cite{Razeghi2018icassp}, and 4) STC for correlated data which are drawn from AR(1) model with the parameter $\rho = 0.5$. We used the same experimental setting as \cite{Razeghi2018icassp}. 
Fig.\ref{fig:DistortionSparsityUnauthorized} shows the reconstruction leakage at a curious server (or an adversary) who knows the encoder and its parameters, but has no knowledge about the correct indices to purify the ambiguated representations. 
The terminology `half ambiguation' is defined as $S_p \!= \! 0.5 (L \! - \!S_x)$, 
and the terminology `full ambiguation' is defined as $S_p \! = \! L - \! S_x$.  
Note that the information security guarantee addressed in \cite{boufounos2011secure}, required keeping the projection parameters secretly. %

\begin{figure*}[!t]
    \centering
     \begin{subfigure}[h]{0.15\textwidth}
        \includegraphics[width=\linewidth, height=2.4cm]{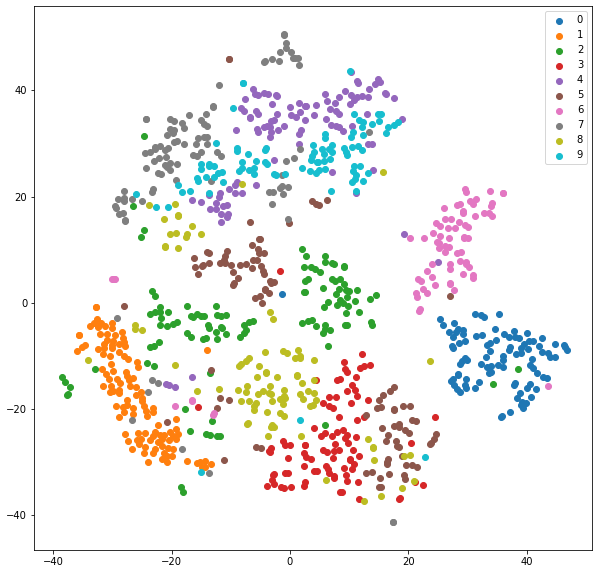}%
        \vspace{-5pt}
         \caption{ }
          \label{fig:tSNE_original}
     \end{subfigure}
     \begin{subfigure}[h]{0.15\textwidth}
        \includegraphics[width=\linewidth, height=2.4cm]{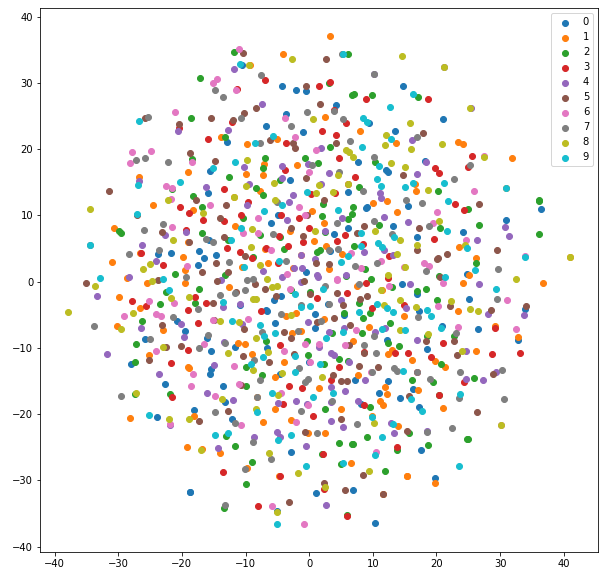}%
        \vspace{-5pt}
         \caption{ }
          \label{fig:tSNE_latent_amb}
     \end{subfigure}
     \begin{subfigure}[h]{0.15\textwidth}
        \includegraphics[width=\linewidth, height=2.4cm]{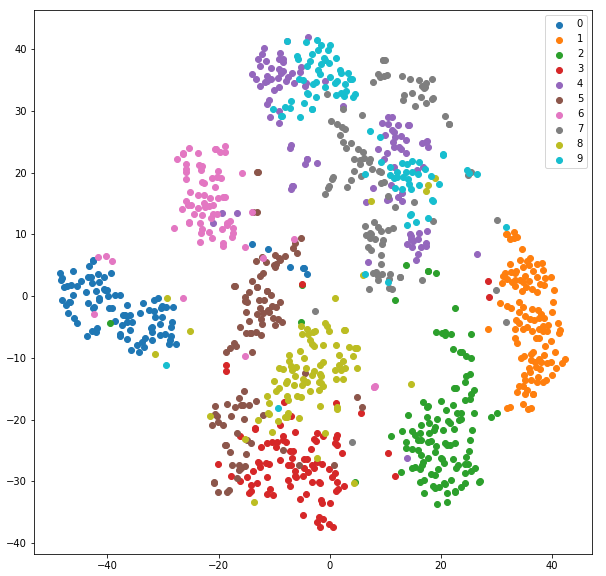}%
        \vspace{-5pt}
         \caption{ }
          \label{fig:tSNE_latent_clean}
     \end{subfigure}
     \begin{subfigure}[h]{0.15\textwidth}
        \includegraphics[width=\linewidth, height=2.4cm]{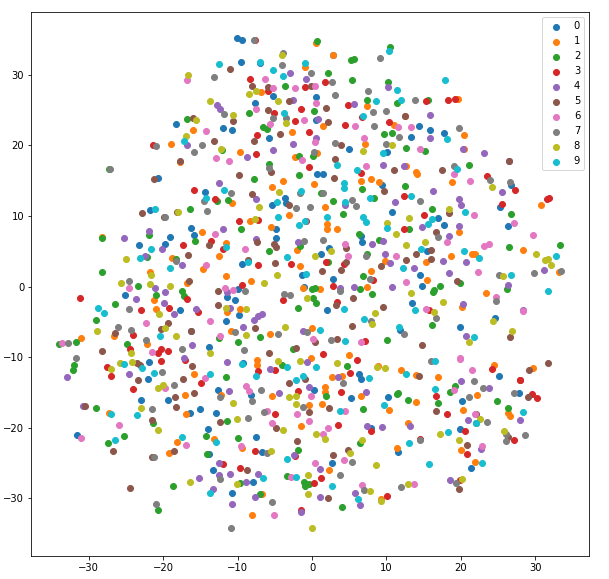}%
        \vspace{-5pt}
         \caption{ }
          \label{fig:tSNE_rec_amb}
     \end{subfigure}  
     \begin{subfigure}[h]{0.155\textwidth}
        \includegraphics[width=\linewidth, height=2.4cm]{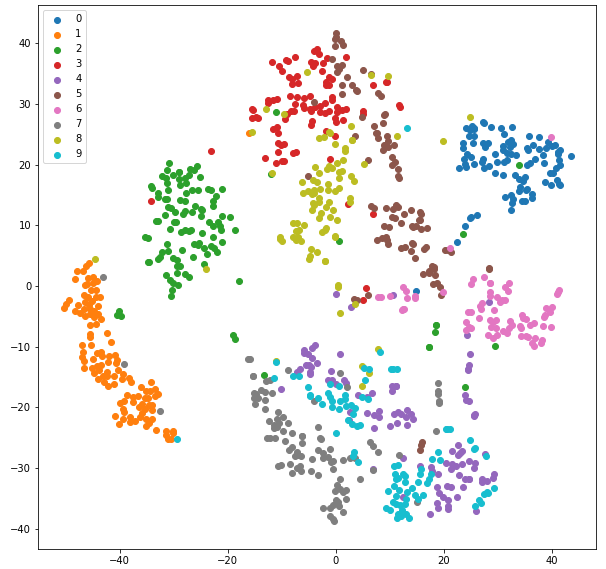}%
        \vspace{-5pt}
         \caption{ }
          \label{fig:tSNE_rec_clean}
     \end{subfigure}   ~~
     \begin{subfigure}[h]{0.15\textwidth}
        \includegraphics[width=\linewidth,, height=2.4cm]{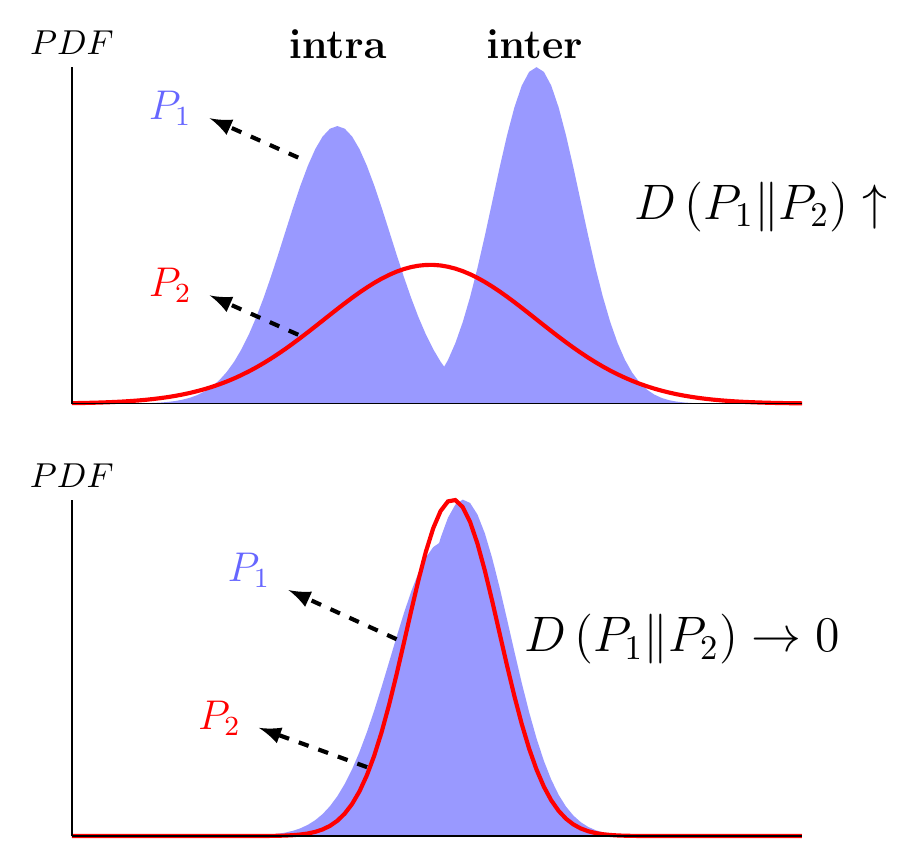}%
        \vspace{-5pt}
         \caption{ }
          \label{fig:KLD}
     \end{subfigure}  
     \caption{t-SNE visualizations from MNIST dataset on: a) original space, b) transformed space with ambiguation, c) transformed space after purification, d) reconstructed space without knowledge of support, e) reconstructed space with knowledge of support; f) conceptual visualization of data clustering leakage.}
     \label{Fig:tSNE}
\end{figure*}

Table~\ref{table:RD} provides a quantitative comparison on reconstructed images using normalized MSE and SSIM (Structural Similarity Index) on MNIST \cite{lecun-mnisthandwrittendigit-2010}, Fashion-MNIST \cite{xiao2017fashion}, and CIFAR-10 \cite{krizhevsky2009learning} databases, where we applied the proposed method on latent representation of a designed convolutional autoencoder in \cite{SohrabBehrooz2020icassp}, setting $L= 128$ and considering one code-map for MNIST and Fashion-MNIST databases and four code-maps for CIFAR-10 database. 
Finally, note that based on these results, our model follows the notion of $\left( \beta, \gamma \right)$-recoverable privacy mechanism, which we defined in Section~\ref{Sec:Preliminaries}. 

As a large-scale retrieval experiment, Figs.~\ref{fig:10_recall_100}-\ref{fig:10_recall_10} depict the recall measure for the CelebA database. The ground-truth was the pixel domain Euclidean distances and the latent code of the network in \cite{SohrabBehrooz2020icassp} is used to measure the approximate distances.


\subsection{Clustering Leakage}


Another potential threat is that the adversary may establish links between the closet disclosed representations. We now discuss database clustering leakage under our model. Note that the proposed mechanism can apply to privacy-preserving clustering applications where the goal is to perform clustering without disclosing the original data. 
The significant benefit of our method is that 
the authorized data users can purify the imposed ambiguation noise. 
However, the adversary will face a combinatorial problem to guesses the correct components. 

Fig.~\ref{Fig:tSNE} provides a qualitative visualization of clustering leakage on MNIST database \cite{lecun-mnisthandwrittendigit-2010}, 
for which t-distributed stochastic neighbor embedding (t-SNE) \cite{maaten2008visualizing} is used to project the underlying space to 2D. 
As illustrated, our model prevent database clustering leakage.  
Denoting by $P_{\mathrm{intra}}$ and $P_{\mathrm{inter}}$ as probability density functions of `intra-cluster' and `inter-cluster' of distances, respectively, Fig.~\ref{fig:KLD}, provides a conceptual visualization of database clustering leakage, where $D\left( P_1 \| P_2 \right) = \mathbb{E}_{P_1}  [ \log \frac{P_1}{P_2}  ]$. 

%
%

%
%



\section{Conclusion}
\label{Sec:Conclusion}


We present a computationally efficient, fairness-aware privacy-preserving nearby search scheme that can be utilized in cloud-based applications. 
The key insight behind our mechanism is that by approximating sparse representation of data points and adding random noise to their orthogonal complement, we can control privacy and utility trade-off in terms of dataset reconstruction and dataset clustering. 
The authorized data users can purify the ambiguated public representation thanks to the knowledge of correct support of the query. 
%







\bibliographystyle{IEEEbib}
\bibliography{strings,refs}

\begin{thebibliography}{10}

\bibitem{wang2017survey}
Jingdong Wang, Ting Zhang, Nicu Sebe, Heng~Tao Shen, et~al.,
\newblock ``A survey on learning to hash,''
\newblock {\em IEEE transactions on pattern analysis and machine intelligence},
  vol. 40, no. 4, pp. 769--790, 2017.

\bibitem{indyk1998approximate}
Piotr Indyk and Rajeev Motwani,
\newblock ``Approximate nearest neighbors: towards removing the curse of
  dimensionality,''
\newblock in {\em Proceedings of the thirtieth annual ACM symposium on Theory
  of computing}. ACM, 1998, pp. 604--613.

\bibitem{gionis1999similarity}
Aristides Gionis, Piotr Indyk, Rajeev Motwani, et~al.,
\newblock ``Similarity search in high dimensions via hashing,''
\newblock in {\em Vldb}, 1999, vol.~99, pp. 518--529.

\bibitem{wang2015learning}
Jun Wang, Wei Liu, Sanjiv Kumar, and Shih-Fu Chang,
\newblock ``Learning to hash for indexing big data—a survey,''
\newblock {\em Proceedings of the IEEE}, vol. 104, no. 1, pp. 34--57, 2015.

\bibitem{datar2004locality}
Mayur Datar, Nicole Immorlica, Piotr Indyk, and Vahab~S Mirrokni,
\newblock ``Locality-sensitive hashing scheme based on p-stable
  distributions,''
\newblock in {\em Annual Symposium on Computational Geometry}. ACM, 2004, pp.
  253--262.

\bibitem{christiani2019fast}
Tobias Christiani,
\newblock ``Fast locality-sensitive hashing frameworks for approximate near
  neighbor search,''
\newblock in {\em Int. Conf. on Similarity Search and Applications}. Springer,
  2019, pp. 3--17.

\bibitem{har2019near}
Sariel Har-Peled and Sepideh Mahabadi,
\newblock ``Near neighbor: Who is the fairest of them all?,''
\newblock in {\em Advances in Neural Information Processing Systems (NeurIPS)},
  2019, pp. 13176--13187.

\bibitem{salakhutdinov2009semantic}
Ruslan Salakhutdinov and Geoffrey Hinton,
\newblock ``Semantic hashing,''
\newblock {\em International Journal of Approximate Reasoning}, vol. 50, no. 7,
  pp. 969--978, 2009.

\bibitem{weiss2009spectral}
Yair Weiss, Antonio Torralba, and Rob Fergus,
\newblock ``Spectral hashing,''
\newblock in {\em Advances in neural information processing systems}, 2009, pp.
  1753--1760.

\bibitem{Sohrab_ISIT2017}
Sohrab Ferdowsi, Slava Voloshynovskiy, Dimche Kostadinov, and Taras Holotyak,
\newblock ``Sparse ternary codes for similarity search have higher coding gain
  than dense binary codes,''
\newblock in {\em IEEE Int. Symp. on Inf. Theory (ISIT)}, 2017.

\bibitem{Razeghi2018icassp}
Behrooz Razeghi and Slava Voloshynovskiy,
\newblock ``Privacy-preserving outsourced media search using secure sparse
  ternary codes,''
\newblock in {\em IEEE Int. Conference on Acoustics, Speech and Signal
  Processing (ICASSP)}, 2018, pp. 1--5.

\bibitem{SohrabBehrooz2020icassp}
Sohrab Ferdowsi, Behrooz Razeghi, Taras Holotyak, Flavio P.~Calmon, and Slava
  Voloshynovskiy,
\newblock ``Privacy-preserving image sharing via sparsifying layers on
  convolutional groups,''
\newblock in {\em IEEE Int. Conference on Acoustics, Speech and Signal
  Processing (ICASSP)}, 2020.

\bibitem{boufounos2011secure}
Petros Boufounos and Shantanu Rane,
\newblock ``Secure binary embeddings for privacy preserving nearest
  neighbors,''
\newblock in {\em IEEE Int. Work. on Inf. Forensics and Security (WIFS)}, 2011,
  pp. 1--6.

\bibitem{weng2016privacy}
Li~Weng, Laurent Amsaleg, and Teddy Furon,
\newblock ``Privacy-preserving outsourced media search,''
\newblock {\em IEEE Transactions on Knowledge and Data Engineering}, vol. 28,
  no. 10, pp. 2738--2751, 2016.

\bibitem{kenthapadi2012privacy}
Krishnaram Kenthapadi, Aleksandra Korolova, Ilya Mironov, and Nina Mishra,
\newblock ``Privacy via the johnson-lindenstrauss transform,''
\newblock {\em arXiv preprint arXiv:1204.2606}, 2012.

\bibitem{rane2013privacy}
Shantanu Rane and Petros~T Boufounos,
\newblock ``Privacy-preserving nearest neighbor methods: Comparing signals
  without revealing them,''
\newblock {\em IEEE Signal Processing Magazine}, vol. 30, no. 2, pp. 18--28,
  2013.

\bibitem{jegou2009burstiness}
Herv{\'e} J{\'e}gou, Matthijs Douze, and Cordelia Schmid,
\newblock ``On the burstiness of visual elements,''
\newblock in {\em IEEE Conf. on Comp. Vision and Pattern Recog. (CVPR)}, 2009,
  pp. 1169--1176.

\bibitem{perronnin2007fisher}
Florent Perronnin and Christopher Dance,
\newblock ``Fisher kernels on visual vocabularies for image categorization,''
\newblock in {\em IEEE Conf. on Comp. Vision and Pattern Recog. (CVPR)}, 2007,
  pp. 1--8.

\bibitem{jegou2010aggregating}
Herv{\'e} J{\'e}gou, Matthijs Douze, Cordelia Schmid, and Patrick P{\'e}rez,
\newblock ``Aggregating local descriptors into a compact image
  representation,''
\newblock in {\em IEEE Conf. on Comp. Vision and Pattern Recog. (CVPR)}, 2010,
  pp. 3304--3311.

\bibitem{babenko2014neural}
Artem Babenko, Anton Slesarev, Alexandr Chigorin, and Victor Lempitsky,
\newblock ``Neural codes for image retrieval,''
\newblock in {\em Europ. \!\! Conf. \!\! on Comp. \!\!\! Vision}. Springer,
  2014, pp. \!584--599.

\bibitem{kingma2014auto}
Diederik~P Kingma and Max Welling,
\newblock ``Auto-encoding variational bayes,''
\newblock in {\em International Conference on Learning Representations (ICLR)},
  2014.

\bibitem{ravishankar2013learning}
Saiprasad Ravishankar and Yoram Bresler,
\newblock ``Learning sparsifying transforms,''
\newblock {\em IEEE Trans. on Signal Processing}, vol. 61, no. 5, pp.
  1072--1086, 2013.

\bibitem{gheisari2020joint}
Marzieh Gheisari, Teddy Furon, and Laurent Amsaleg,
\newblock ``Joint learning of assignment and representation for biometric group
  membership,''
\newblock in {\em ICASSP 2020-2020 IEEE International Conference on Acoustics,
  Speech and Signal Processing (ICASSP)}. IEEE, 2020, pp. 2922--2926.

\bibitem{gheisari2019group}
Marzieh Gheisari, Teddy Furon, and Laurent Amsaleg,
\newblock ``Group membership verification with privacy: Sparse or dense?,''
\newblock in {\em 2019 IEEE International Workshop on Information Forensics and
  Security (WIFS)}. IEEE, 2019, pp. 1--7.

\bibitem{gheisari2019aggregation}
Marzieh Gheisari, Teddy Furon, Laurent Amsaleg, Behrooz Razeghi, and Slava
  Voloshynovskiy,
\newblock ``Aggregation and embedding for group membership verification,''
\newblock in {\em ICASSP 2019-2019 IEEE International Conference on Acoustics,
  Speech and Signal Processing (ICASSP)}. IEEE, 2019, pp. 2592--2596.

\bibitem{Kostadinov2018:EUVIP}
Dimche Kostadinov, Slava Voloshynovskiy, and Sohrab Ferdowsi,
\newblock ``Learning overcomplete and sparsifying transform with approximate
  and exact closed form solutions,''
\newblock in {\em European Workshop on Visual Information Processing}, 2018.

\bibitem{Razeghi2017wifs}
Behrooz Razeghi, Slava Voloshynovskiy, Dimche Kostadinov, and Olga Taran,
\newblock ``Privacy preserving identification using sparse approximation with
  ambiguization,''
\newblock in {\em IEEE Int. Work. on Info. Forensics and Security (WIFS)},
  2017, pp. 1--6.

\bibitem{lecun-mnisthandwrittendigit-2010}
Yann LeCun and Corinna Cortes,
\newblock ``{MNIST} handwritten digit database,''
\newblock 2010.

\bibitem{xiao2017fashion}
Han Xiao, Kashif Rasul, and Roland Vollgraf,
\newblock ``Fashion-mnist: a novel image dataset for benchmarking machine
  learning algorithms,''
\newblock {\em arXiv preprint arXiv:1708.07747}, 2017.

\bibitem{krizhevsky2009learning}
Alex Krizhevsky, Geoffrey Hinton, et~al.,
\newblock ``Learning multiple layers of features from tiny images,''
\newblock 2009.

\bibitem{maaten2008visualizing}
Laurens van~der Maaten and Geoffrey Hinton,
\newblock ``Visualizing data using t-sne,''
\newblock {\em Journal of machine learning research}, vol. 9, no. Nov, pp.
  2579--2605, 2008.

\end{thebibliography}

\end{document}